\documentclass[runningheads]{cl2emult}

\usepackage{graphicx} % standard LaTeX graphics tool
\usepackage{subeqnar} % subnumbers individual equations
\usepackage{multicol} % used for the two-column index
\usepackage{cropmark} % cropmarks for pages without
\usepackage{eso}      % placeholder for figures
\begin{document}
\title*{Deep Near--Infrared Imaging Surveys 
and the Stellar Content of High Redshift Galaxies}
\toctitle{Deep Near--Infrared Imaging Surveys}
\titlerunning{Deep Near--Infrared Imaging Surveys}
\author{Mark Dickinson\inst{1}
\and Casey Papovich\inst{1,2}
\and Henry C.\ Ferguson\inst{1}
}
\authorrunning{Mark Dickinson et al.}
\institute{Space Telescope Science Institute, Baltimore MD 21218, USA
\and The Johns Hopkins University, Baltimore MD 21218, USA}

\maketitle              % typesets the title of the contribution

\begin{abstract}
Deep, near--infrared imaging surveys have been motivated 
by the desire to study the rest--frame optical properties 
and stellar content of galaxies at high redshift.  Here we briefly 
review their history, and illustrate one application, using {\it HST} 
NICMOS imaging of the Hubble Deep Field North to examine the rest--frame 
optical luminosities and colors of galaxies at $2 < z < 3$, and to 
constrain their stellar masses.  The rest--frame $B$--band luminosity 
density at $z \approx 2.5$ is similar to that in the local universe, 
but the galaxies are evidently less massive, with rapid star formation 
and low mass--to--light ratios.  There are few candidates for red, 
non--star--forming galaxies at these redshifts to the HDF/NICMOS 
limits.  We estimate a stellar mass density at $2 < z < 3$ that is 
$\sim 5$\% of the present--day value, with an upper bound of $\leq 30$\%.
Future headway will come from wide--field ground--based surveys, 
multiplexing infrared spectrographs, and new space--based 
facilities such as {\it HST}/WFC3 and {\it SIRTF}/IRAC.
%\index{abstract}
\end{abstract}

\section{Introduction and history}

Observing proposals and article introductions almost universally
list a set of basic, interrelated themes that motivate deep 
near--infrared (NIR, here regarded as 1--3~$\mu$m) blank 
sky surveys:
\begin{itemize}
\item The integrated stellar spectra of normal galaxies peak in the NIR
\item The stellar component of the extragalactic background peaks in the NIR
\item NIR light measures familiar rest--frame optical wavelengths at high $z$
\item Optical/NIR rest--frame light comes primarily from mid-- to low--mass stars
	with long lifetimes relative to $H_0^{-1}$
\item NIR light traces total stellar content/mass
\item Evolutionary corrections are smaller and easier to model
\item $k$--corrections are small or negative and similar for most galaxy types
\item Effects of dust extinction are greatly reduced
\item Access to $z > 6$, where galaxy light shifts beyond optical wavelengths.
\end{itemize}
The earliest, heroic efforts~\cite{Boughn86,Collins88} used 
single--element photometers to search for sky background fluctuations
from primeval galaxies (PGs) whose light might be redshifted beyond 
the optical wavelength range.   The field really came to life with the 
advent of array detectors, leading to the first faint NIR imaging 
surveys~\cite{Elston88,Cowie88}.  It is interesting to read these 
and other early papers, where we find discoveries, concerns and 
hypotheses that have stayed with us ever since: PGs, extremely red objects 
(``EROs'' as PG candidates, or high--$z$ ellipticals, or dust--enshrouded 
galaxies), extremely blue objects (with rapid, cosmologically significant
star formation), ERO clustering, UV--excess ellipticals, 
photometric redshifts, NIR number counts (to constrain space 
curvature and/or galaxy evolution), etc.  While many of the
issues remain the same today, the data quality has advanced tremendously,
largely driven by progress in array technology, and most recently 
by the leap into space with {\it HST}/NICMOS (affording high angular 
resolution and far lower backgrounds).  Survey sensitivities have 
improved by a factor of $\sim 1000$, source densities have increased 
$\sim 200$--fold, and we can now image the detailed morphologies
of high redshift galaxies in their rest--frame optical light.

\section{Infrared observations of the Hubble Deep Fields}

The {\it HST} WFPC2 and STIS observations of the Hubble Deep Fields 
(HDFs, North and South) are the deepest optical images of the sky,
and correspondingly deep NIR observations of these areas 
are valuable for all the reasons outlined above. The HDF--N was 
observed from the ground in several different NIR 
programs~\cite{Hogg97,Barger98,Dickinson98,Hogg00a,Hogg00b}, 
while the HDF--S has been imaged from the ESO NTT~\cite{daCosta98} 
and more recently to with ISAAC on the VLT~\cite{Franx00}.  
The deepest observations at 1.1--1.6~$\mu$m have come from 
{\it HST} NICMOS imaging of the HDF--N~\cite{Thompson99,Dickinson99} 
and for the HDF--S NICMOS field~\cite{Fruchter01} (which is distinct 
from the HDF--S WFPC2 field).  The depth and angular resolution of 
our ``wide--field'' (only $\sim 6$~arcmin$^2$, smaller than the 
first NIR array surveys!) HDF--N/NICMOS program, combined with the 
great wealth of supporting imaging and spectroscopy at other wavelengths 
from the ground and from space, make this a premier resource for studying 
the NIR properties of galaxies at high redshift.  Discussion of the NIR
morphological and photometric properties of galaxies at 
$0 < z < 2$~\cite{Dickinson00a} and $2 < z < 3.5$~\cite{Dickinson00b},
and of galaxy candidates at 
$z\gg4$~\cite{Dickinson00b,Dickinson00,Lanzetta98,Lanzetta99,Yahata00,Thompson01} 
have appeared elsewhere.

\section{Stellar populations of galaxies at $2 < z < 3$}

\begin{figure}
\centering
\includegraphics[width=\textwidth]{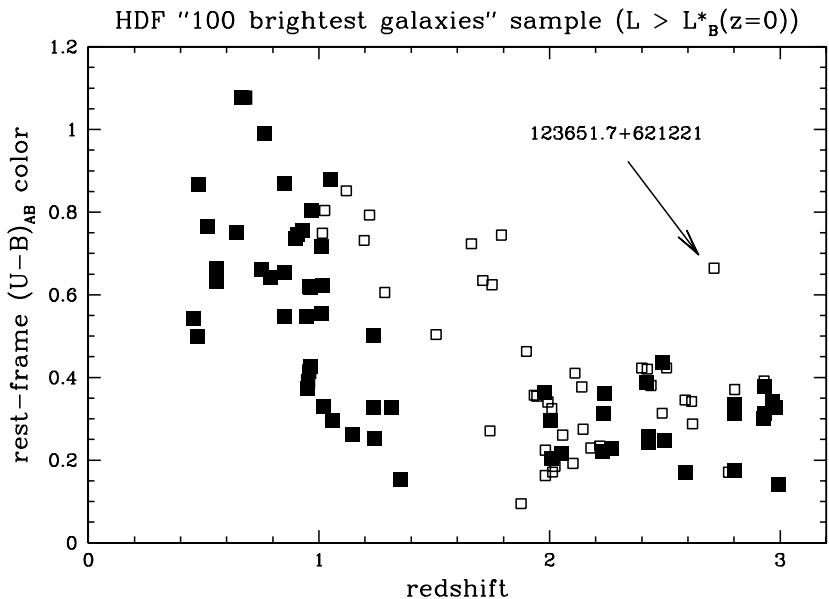}
\caption[]{Rest--frame $U-B$ color vs.\ redshift for HDF--N
galaxies with rest--frame $M_B < -20.3$ ($\Omega_M = 0.3$, 
$\Omega_\Lambda = 0.7$, $h=0.7$).  Filled and open symbols 
indicate galaxies with spectroscopic and photometric redshifts,
respectively.  There is a strong color trend with redshift;
nearly all galaxies at $2 < z < 3$ are bluer than comparably 
luminous galaxies at $z < 1$.  One redder object with
$z_{\rm phot} = 2.7$ is marked;  this object is also 
a radio~\cite{Richards99} and X--ray source~\cite{Hornschemeier00}, 
and may be detected at 15~$\mu$m~\cite{Aussel99} and 
1.3~mm~\cite{Downes99} as well.}
\label{UmBvsz}
\end{figure}

We have carried out a detailed study~\cite{Papovich01} of the 
stellar population properties of star--forming ``Lyman break galaxies'' 
(LBGs) from the HDF--N at $2 < z < 3.5$, using NICMOS data to extend 
previous work based on ground--based NIR photometry~\cite{Sawicki98}.  
Using a sample of 33 spectroscopically confirmed galaxies~\cite{Cohen00}, 
we compared 7--band (0.3--2.2~$\mu$m, observed frame) photometry to empirical 
spectral templates for nearby galaxies and to population synthesis models 
in order to evaluate constraints on the LBGs' stellar content and 
evolutionary histories.  The LBGs are much bluer than local, Hubble 
sequence galaxies, and than comparably luminous HDF galaxies at lower
redshift (Fig.~\ref{UmBvsz}), but are similar to nearby, UV--bright 
starburst galaxies~\cite{Dickinson00b,Papovich01}.  

\begin{figure}
\centering
\includegraphics[width=\textwidth]{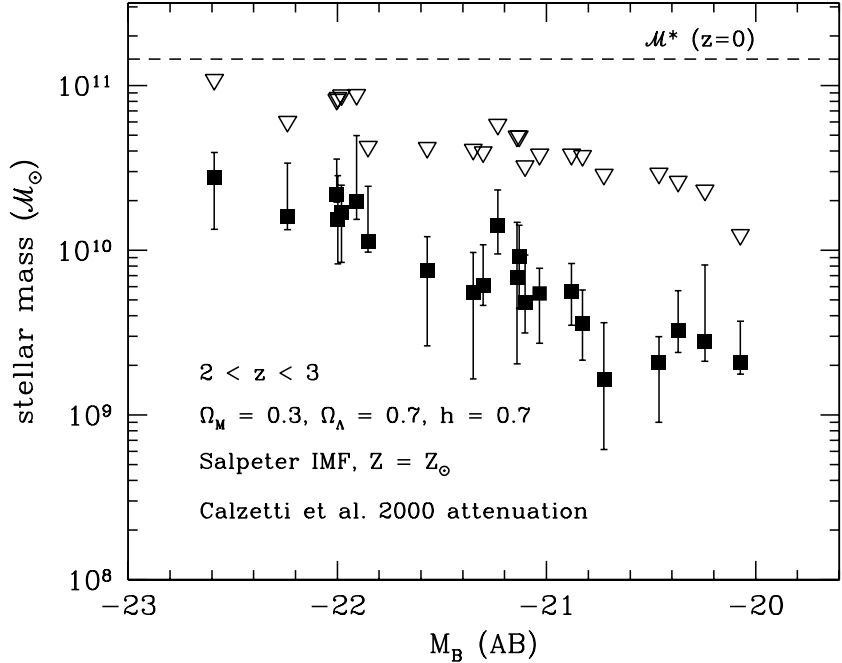}
\caption[]{Stellar masses for HDF--N Lyman break galaxies derived
from stellar population model fitting~\cite{Papovich01}.  The filled
points show best--fitting mass estimates using solar metallicity, 
Salpeter IMF models, and assuming exponential star formation histories, 
while the error bars show 68\% confidence intervals.  The downward
triangles are upper mass limits allowing for the presence of 
an underlying, maximum $\mathcal{M}/L$ stellar population formed 
at $z = \infty$.  The dashed line shows the present--day 
characteristic $\mathcal{M}^\ast$ stellar mass~\cite{Cole00} 
for the same IMF.}
\label{massplot}
\end{figure}

Even with high--quality {\it HST} optical/infrared photometry,
we find only weak constraints on most parameters of the LBG
stellar populations, with degeneracies between age, star formation
time scale, metallicity and extinction.  Perhaps the best constraints, 
however, are those on the total stellar mass $\mathcal{M}$ 
(see Fig.~\ref{massplot}).  
If the LBG star formation history $\Psi(t)$ is modeled by an 
exponential, $\Psi \propto e^{-t/\tau}$, with $t$ and $\tau$ 
(and extinction) as free parameters, then with fixed assumptions
about the IMF and metallicity, the typical 68\% confidence 
interval on log~$\mathcal{M}$ is approximately $\pm 0.25$~dex.
For LBGs with $L^\ast$ UV luminosities~\cite{Steidel99}, the 
inferred stellar masses (assuming a Salpeter IMF, and varying
the model metallicities) are 1 to $2\times 10^{10} \mathcal{M}_\odot$ 
for $\Omega_M = 0.3$, $\Omega_\Lambda = 0.7$, $h = 0.7$ (used here 
unless otherwise noted).  These are $\sim 1/10$th the stellar 
masses of $L^\ast$ galaxies today~\cite{Cole00}.  We may compare 
these masses to those derived from virial estimates using nebular
line--widths and {\it HST}--measured sizes~\cite{Pettini01a,Pettini01b},
which are also $\sim 10^{10} \mathcal{M}_\odot$.  This suggests
that these kinematic measurements underestimate the full mass 
of the dark matter halo, reinforcing a point emphasized by 
Max Pettini in this volume.

We set an upper bound on the allowable stellar mass by considering 
how much light from a hypothetical, maximally old stellar population 
(formed at $z = \infty$) could be hidden beneath the glare of the 
young, star--forming population.  On average, this upper bound
is a factor of $\sim 5$ to $6\times$ larger than the mass derived 
for the ``young'' models.  If this were generally the case, however, 
then virtually all galaxies with previous generations of star formation
at $z \gg 3$ must {\it also} be forming stars rapidly at $2 < z < 3$.
We see very few candidates for mature, red, non--star--forming galaxies 
in this redshift range, even with a NICMOS--selected sample where 
photometric redshifts should, in principle, readily identify
such galaxies if they are present (see Fig.~\ref{UmBvsz}).

\section{The optical luminosity function and stellar mass 
density at $2 < z < 3$}

\begin{figure}
\centering
\includegraphics[width=\textwidth]{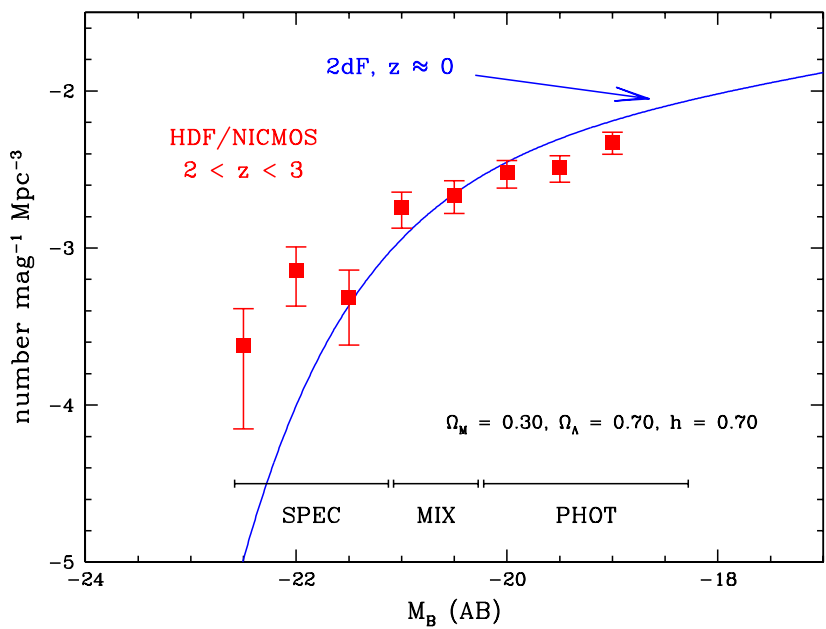}
\caption[]{Rest--frame $B$--band luminosity distribution 
for galaxies at $2 < z < 3$ in the HDF--N.  Horizontal bars 
indicate magnitude ranges where spectroscopic or photometric 
redshifts, or a mix of the two, have been used.  The data
have not been corrected for incompleteness, which may affect 
the fainter points.  Error bars indicate Poisson uncertainties
only.  The local luminosity function from the 2dF 
survey~\cite{Folkes99} is shown for comparison.}
\label{lfBrest}
\end{figure}

Using photometric and spectroscopic redshifts and NICMOS photometry
for HDF--N galaxies, we may examine the rest--frame $B$--band 
luminosity distribution of galaxies at $2 < z < 3$ (Fig.~\ref{lfBrest}).
Over the luminosity range we can examine, this is not dissimilar
to the local $B$--band luminosity function (LF)~\cite{Folkes99}.
Galaxies with $2 < z_{\rm phot} < 3$ and $M_B < -18.75$ contribute
a total blue luminosity density
$\rho_B = 5.2 \times 10^{26}\ \mathrm{erg~s^{-1}~Mpc^{-3}}$.
Without further correction for incompleteness or extrapolation
to fainter magnitudes, this is $\sim 1.5\times$ that from
the integrated 2dF LF, and nearly equal to that from the preliminary
SDSS LF~\cite{Blanton00}.

Although the optical luminosity densities are similar at $z=0$ and $z=3$, 
the implied stellar mass densities are quite different.  Galaxies at 
$2 < z < 3$ are far bluer than local counterparts (Fig.~\ref{UmBvsz}),
indicating much smaller mass--to--light ratios.  From our 
modeling~\cite{Papovich01} for 22 HDF--N LBGs at $2 < z < 3$, 
we derive an average $\langle \mathcal{M}/L_B \rangle = 0.10$ to 0.25
(in solar units), depending on assumptions about metallicity and IMF.  
These values refer to {\it emergent} luminosities, i.e., here $L_B$ is 
{\it not} corrected for the effects of extinction -- the {\it intrinsic} 
$\mathcal{M}/L$ for the stellar populations are smaller.  There is a trend 
of $\mathcal{M}/L$ with rest--frame color, amounting to a factor of $\sim 2$ 
over the observed color range of the spectroscopic LBG sample,
but we neglect this here and apply $\langle \mathcal{M}/L_B \rangle$ 
from the spectroscopic LBGs to the photometric redshift sample.  
The majority of galaxies with photometric redshifts $2 < z_{\rm phot} < 3$ 
have colors similar to or bluer than those in the spectroscopic 
sample (Fig.~\ref{UmBvsz}).

Restricting our attention to a Salpeter IMF, the starburst dust 
attenuation law~\cite{Calzetti00}, and model metallicities
0.2 to 1$\times Z_\odot$, we estimate a stellar mass density 
$\rho_* = 1.7$ to $2.9 \times 10^7 \mathcal{M}_\odot \mathrm{Mpc}^{-3}$ 
at these redshifts.  Comparing this to the present--day stellar mass density 
computed from the 2dF+2MASS $K$--band luminosity function~\cite{Cole00}, 
using the same IMF assumptions, we find $\rho_*(z=2.5) / \rho_*(z=0) = 0.034$ 
to 0.058.  It is also 8--14$\times$ smaller than the estimated
mass density from bright galaxies at $z \approx 0.9$~\cite{Brinchmann00}.
This is presumably a lower limit for several reasons.  
First, our assumed cosmology results in a nearly minimal luminosity 
density for currently acceptable values of the cosmological parameters.
An Einstein--de~Sitter model increases by the luminosity and mass
densities by $\sim$80\%. Second, we have 
made no corrections for incompleteness in the NICMOS--selected 
galaxy sample, nor any attempt to extrapolate to objects fainter than 
the HDF/NICMOS detection limit.  Finally, as described in \S 2, the 
galaxy masses may be larger if there were earlier generations of 
star formation.  If we assign {\it every} LBG its {\it maximum} 
(at 68\% confidence) stellar mass (see \S 2) allowing for 
a older generation of stars formed at $z = \infty$, then for the 
adopted cosmology we may set a conservative upper bound on the total 
stellar mass density contained within NICMOS--detected galaxies, 
$\rho_*(z=2.5) / \rho_*(z=0) < 0.30$.

This upper bound is barely consistent with the hypothesis that
all stars in present--day galactic spheroids formed at 
$z > 3$~\cite{Renzini99}.  In this scenario, all galaxies must
have already formed most of their stars at $z \gg 3$, but must
also be forming more stars at $z \approx 2.5$ (since there are
are few candidates for evolved, non--star--forming HDF galaxies
at that redshift).  As described above, a more direct accounting 
for the mass present in NICMOS--detected galaxies at $z \approx 2.5$ 
implies a much smaller fraction, $\sim 5$\%, of the present--day 
stellar mass density.

\section{Future directions}

Infrared surveys described here have really just begun to address
the most important questions about the mass assembly history of 
galaxies.  A new generation of infrared instrumentation, on the 
ground and in space, will carry us further along down this road.
Deep NIR surveys are still limited to very small solid angles
and thus volumes at high redshift.  New large format detectors
will greatly improve this situation, although unfortunately
very few cameras for 8--10m telescopes (whose aperture is 
really needed to study $z > 2$ galaxies) are being configured
with wide fields of view.  The first multi--object NIR spectrographs
are just now coming on line;  these will permit wholesale spectroscopy
of distant galaxies at rest--frame optical wavelengths, offering
a means of measuring kinematic masses and chemical abundances at
high redshift.  On {\it HST}, the infrared channel of Wide Field Camera~3 
will offer a big advance for faint galaxy surveys, offering a field
of view $5.5\times$ larger than that of NICMOS, with 1.7$\times$ better 
pixel sampling.  Deep 1.0--1.6~$\mu$m observations of HDF--sized regions 
will become routine, enabling wholesale studies of the rest--frame
optical morphologies, luminosities and colors of galaxies at $z < 3$,
and color--selected surveys for galaxies at $z > 6$.

Despite the promises usually made for NIR surveys (see \S 1),
our census of the stellar content of galaxies at $z > 2$ is still
fundamentally limited by the wavelengths at which we can observe.
The $H$ and $K$--bands measure rest--frame $B$ and $V$--band light 
at $z = 3$, leaving large uncertainties on estimates of stellar mass, 
ages, and other such parameters, while brave first attempts at 3--7~$\mu$m
do not go deep enough to detect galaxies at $z > 2$~\cite{Serjeant97,Hogg00b}.
The {\it SIRTF} Infrared Array Camera (IRAC), observing at
3.6--8.0~$\mu$m, can measure rest--frame $K$--band light from 
galaxies out to $z \approx 3$ and $\lambda_0 > 1$~$\mu$m emission 
out to $z = 7$.   Extremely deep exposures will be required, 
however, to detect ordinary galaxies at such large redshifts.   
We will be carrying out a {\it SIRTF} Legacy Program, 
the Great Observatories Origins Deep Survey (GOODS), which will 
push observations at 3.6--24~$\mu$m to their limits in two 
fields (the HDF--N and Chandra Deep Field South) totaling
approximately 330~arcmin$^2$.   The survey goal is to provide 
multiwavelength data suitable for tracing the mass assembly 
history of galaxies and their energetic output from star formation 
and AGN activity out to the highest accessible redshifts.  
The {\it SIRTF} data, along with extensive supporting 
observations from ESO and other facilities, will be distributed 
to the community, providing a rich archive for research 
and a pathfinder to future work with NGST.\\

MD would like to thank the conference organizers for hosting
this important and timely meeting.  This work was supported
by NASA grant GO-07817.01-96A.

%INDEX%%%%%%%%%%%%%%%%%%%%%%%%%%%%%%%%%%%%%%%%%%%%%%%%%%%%%%%%%%%%%%%
%\clearpage
%\addcontentsline{toc}{section}{Index}
%\flushbottom
%\printindex
%%%%%%%%%%%%%%%%%%%%%%%%%%%%%%%%%%%%%%%%%%%%%%%%%%%%%%%%%%%%%%%%%%%%%

\end{document}